\documentclass{elsart3p}

\usepackage{amssymb}
\usepackage{graphicx}

\begin{document}

\begin{frontmatter}



\title{Robustness of $s$-wave superconductivity against Coulomb interactions in Na$_x$CoO$_2$}


\author[a]{Keiji Yada}
\author[a]{Hiroshi Kontani}

\address[a]{Department of Physics, Nagoya University, Furo-cho, Chikusa-ku, Nagoya, 464-8602, Japan}

\begin{abstract}
We study the depairing effect due to Coulomb interactions in Na$_x$CoO$_2$.
We consider the electron-phonon coupling and the Coulomb interactions,
and determine $T_c$ for $s$-wave superconductivity by solving the linearized Eliashberg equation.
When we consider shear phonons as well as breathing phonons,
$T_c$ is enhanced by Suhl-Kondo (SK) mechanism.
Since SK mechanism is strong against Coulomb interactions,
$T_c$ remains finite even if strong Coulomb interactions cancel out the attractive force due to breathing phonons.
The orbital degree of freedom is important to understand the mechanism of superconductivity in Na$_x$CoO$_2$.

\end{abstract}

\end{frontmatter}

%
%
%
%

In Na$_x$CoO$_2$, although several LDA calculations suggest that there are hole pockets composed of $e_g'$ orbitals.
However, the top of the $e_g'$ bands locate below (but very close to) the Fermi level in ARPES measurements.
Theoretically, we have shown that a weak pseudogap behavior observed in susceptibility or density of states (DOS)
can be explained only when $e_g'$ hole pockets are absent\cite{label1}.
The closeness of the top of $e_g'$ bands to the Fermi level suggests that
the orbital degree of freedom may play a important role in the superconductivity of Na$_x$CoO$_2$.
We have shown that relatively high $s$-wave $T_{\rm c}$ is realized in Na$_x$CoO$_2$
in case we consider $E_{1g}$ phonons (shear phonons) as well as $A_{1g}$ phonons (breathing phonons)
since the interband transition of Cooper pairs are induced by shear phonons\cite{label2}.
This enhancement of $T_{\rm c}$ due to shear phonons is understood as Suhl-Kondo (SK) mechanism
where superconductivity is realized by the interband hopping of the cooper pairs.
Even if the $e_g'$ bands are valence bands,
this mechanism is valid if the top of the $e_g'$ bands locates near Fermi level.
We call this mechanism of superconductivity the valence-band SK mechanism.
Thus, we inferred that Na$_x$CoO$_2$ is an $s$-wave superconductor thanks to the orbital degree of freedom.
However, it is not clear whether $s$-wave superconductivity can be realized in Na$_x$CoO$_2$
since the strong Coulomb interactions would reduce $T_{\rm c}$.

In this paper,
we study the depairing effect due to Coulomb interactions
in a multi-orbital system which describes the effective bands of Na$_x$CoO$_2$.
In this model, we consider the direct Coulomb repulsion $U$ and the pair hopping $J$.
The depairing effect on the effective attractive force between electrons in the same orbital
is given by $U$,
and that for the interband transition of Cooper pairs due to shear phonons
is caused by $J$.
We calculate the reduction of $T_{\rm c}$,
and find that the SK effect due to the shear phonon is strong against Coulomb interactions.

\begin{figure}
\begin{center}
\includegraphics[scale=0.35]{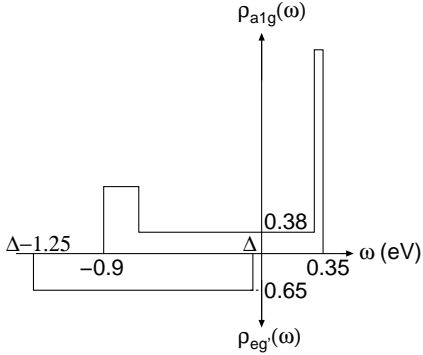}
\caption{DOS of Na$_x$CoO$_2$. The unit of energy is eV.}\label{dos}
\end{center}
\end{figure}

The electron-phonon coupling between $t_{2g}$ ($a_{1g}+e_g'$) electrons
and relevant optical phonon (breathing and shear phonons) are represented by four coupling constant
$a_1$, $a_2$, $b_1$, $b_2$ which is shown elsewhere\cite{label2}.
We calculate $T_{\rm c}$ by solving the linearized Eliashberg equation as follows.
\begin{eqnarray}
&&\lambda \hat\phi({\rm i}\varepsilon_n)=T\sum_{{\rm i}\varepsilon_n}\int_{-\infty}^\infty\!\!\! d\omega \frac{\hat z(\omega) \hat\rho(\omega)}{\varepsilon_{n'}^2+\omega^2}\hat V({\rm i}\varepsilon_n-{\rm i}\varepsilon_{n'})\hat\phi({\rm i}\varepsilon_{n'}),\nonumber\\&& \label{eliash}\\ 
&&\hat V({\rm i}\omega_n)\!=
\!\!\left(
\begin{array}{cc}
a_1^2&2b_1^2\\
b_1^2&a_2^2+2b_2^2
\end{array}
\right)
\!\!D({\rm i}\omega_n)\!-\!\!
\left(
\begin{array}{cc}
U&2J\\
J&U+J
\end{array}
\right),\label{v}
\end{eqnarray}
where $\hat\phi({\rm i}\varepsilon_n)=(\phi_{a1g}({\rm i}\varepsilon_n),\phi_{eg'}({\rm i}\varepsilon_n))$
is a column vector of gap functions.
$D({\rm i}\omega_n)=2\omega_{\rm D}/(\omega_{\rm D}^2+\omega_n^2)$ is the phonon's Green function, where $\omega_{\rm D}$ is the Debye frequency of phonon.
We put $\omega_{\rm D}=550$ (cm$^{-1}$) for breathing and shear phonons.
$\varepsilon_n=(2n-1)\pi T$ and $\omega_n=2n\pi T$ are Matsubara frequencies.
We show the expression of the renormalization factor $z(\omega)$ for $|\omega|<\omega_D$ elsewhere\cite{label3}.
We put $z(\omega)=1$ for $|\omega|>\omega_D$ for simplicity
since the energy range where the band structure is renormalized is $|\omega|<\omega_D$.
The DOS which we use in this calculation is shown in Fig. \ref{dos}.
This reproduces the bandwidth and the DOS near the Fermi level which is obtained by lattice model\cite{label1}.
For $a_{1g}$ band, we add the excess DOS in the band edge
to reproduce the total number of $d$-electron and overall shape roughly.
Note that the integration of DOS does not reach 1
because the ratio of the number of the $d$-electron in three $t_{2g}$ bands near Fermi level is about 0.8
in the $d$-$p$ model.
$T_{\rm c}$ is determined when $\lambda$ in left-hand side of Eq. \ref{eliash} is  equal to 1.

\begin{figure}
\begin{center}
\includegraphics[scale=0.45]{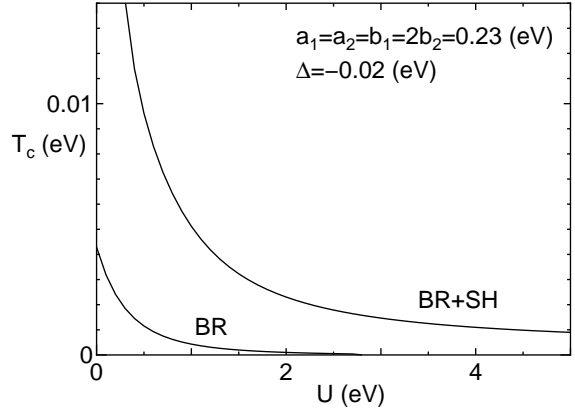}
\caption{$U$-dependence of $T_{\rm c}$ for $J=U/10$.}\label{tc}
\end{center}
\end{figure}

Fig. \ref{tc} shows the $U$-dependence of $T_{\rm c}$ for $J=U/10$,
$\Delta=-0.02$ (eV) and $a_1=a_2=b_1=2b_2=0.23$ (eV).
When we take account of only breathing phonons, $T_{\rm c}$ reduces rapidly and it  vanishes at $U=U_{\rm c}=2.8$ (eV).
On the other hand, if we consider the shear phonons as well as breathing phonons,
$T_{\rm c}$ remains finite even if $U$ exceed $U_{\rm c}$.
This is because there is the effective attractive force due to interband hopping induced by shear phonons.
This interaction is seldom weakened by Coulomb interactions since $J\ll U$.
Thus, the $s$-wave superconductivity with $T_{\rm c}\sim5$K can be realized by this mechanism
even if we consider the strong Coulomb interactions.
In this calculation, the renormalization of $U$ and $J$ due to retardation is included automatically
because the attractive force due to electron-phonon coupling depend on Matsubara frequency
while repulsive force due to Coulomb interactions does not.
We derive the renormalized $U^*$ and $J^*$ elsewhere\cite{label3};
it is shown that the renormalization for $J$ is larger than that for $U$.
Therefore, interband transition due to shear phonon is not prohibited by the Coulomb interactions.

In conclusion, we studied the depairing effect due to Coulomb interactions in Na$_x$CoO$_2$,
and found that $T_{\rm c}$ remains finite even in the presence of the strong Coulomb interactions
when we consider both breathing and shear phonons,
while $T_{\rm c}$ vanishes rapidly when we consider breathing phonons only.
Even if the direct Coulomb repulsion cancels out the attractive force induced by breathing phonons,
the $s$-wave superconductivity due to SK mechanism would be realized.
Thus, the superconductivity due to SK mechanism is realized
even if the $e_g'$ bands are valence bands.

\end{document}